# Monte-Carlo simulation of supercooled liquids using a self-consistent local temperature

Ralph V. Chamberlin and Kurt J. Stangel

Department of Physics and Astronomy, Arizona State University, Tempe, AZ 85287-1504

Abstract

We combine Creutz energy conservation with Kawasaki spin exchange to simulate the microcanonical dynamics of a system of interacting particles. Relaxation occurs via Glauber spin-flip activation using a self-consistent temperature. Heterogeneity in the dynamics comes from finite-size constraints on the spin exchange that yield a distribution of correlated regions. The simulation produces a high-frequency response that can be identified with the boson peak, and a lower-frequency peak that contains non-Debye relaxation and non-Arrhenius activation, similar to the primary response of supercooled liquids.

PACS numbers: 64.70.Pf, 61.20.Lc, 65.20.+w, 87.53.Wz

Key words: nanothermodynamics, supercooled liquids, Monte Carlo, microcanonical

Computer simulations can provide unique insight into the thermal and dynamic properties of complex systems [1]. The two main categories of simulation techniques are molecular dynamics (MD), which often involves fundamental physical principles such as conservation of energy and conservation of momentum; and Monte Carlo (MC) which is usually optimized for computational efficiency. For maximum efficiency we use the Ising model for binary degrees of freedom ("spins") that may be up (+1) or down (-1), but key concepts from the model can be applied to other systems, such as molecular rotation in supercooled liquids. Although some MC algorithms have incorporated realistic constraints [2], including Kawasaki spin exchange [3] where the net alignment (angular momentum) is conserved, most simulations utilize a fixed-$T$ ensemble where each spin is assumed to be in direct contact with an infinite external heat bath. A notable exception is Creutz' [4,5] simulation of the Ising model, where internal energy is conserved by adding conjugate variables that act as a reservoir of kinetic energy. The primary step in our kinetic-Creutz (k-C) model is to combine Creutz and Kawasaki dynamics, yielding fully-closed (microcanonical) constraints, appropriate for fast degrees of freedom [6,7]. Moreover, the internal reservoir allows the spin system to be isolated from external sources of heat, so that the surrounding sample provides a self-consistent thermal bath, and the combined entropy of the spin system and its bath can be maximized to find the true thermal equilibrium.

Dynamical constraints alter the thermal fluctuations and equilibrium properties of finite systems [8]. If the system is homogeneous, fluctuation effects can be removed by extrapolating to infinite size [9], but several techniques have shown that viscous liquids contain internal fluctuations that are dynamically heterogeneous [1,10,11,12]. Thus the response involves nanoscale regions that are effectively uncorrelated with neighboring regions, so that the fluctuations form an ensemble of independent small systems. Although small-system thermodynamics was originally developed for isolated small systems, an ensemble of independent fluctuations should also be treated using this "nanothermodynamics" [13,14]. Subdividing a bulk sample into an ensemble of small regions has long been a theoretical tool [15], but here we give the regions a physical identity by relating them to the length scale for correlated spin exchange. Most MC algorithms impose homogeneous dynamics by utilizing a fixed-$T$ ensemble, explaining why previous simulations showing heterogeneity have used MD [16]. The local source of kinetic energy in the Creutz model allows dynamical heterogeneity in MC simulations.

MC dynamics with several realistic features is facilitated by the Creutz Hamiltonian $H_C = K+U$. Here the potential energy is the usual Ising term $U = -\frac{1}{2} J \Sigma_{<ij>} \sigma_i \sigma_j$, where $\sigma_i = \pm 1$ is the alignment of each spin, $J>0$ is the interaction energy, and the sum is over all pairs of nearest-neighbor spins. The kinetic energy term is $K = 2J \Sigma_i k_i$, where the conjugate variable for $\sigma_i$ is an integer that must be $k_i \geq 0$ if it is to function as a kinetic energy. One realistic feature is that both the spin system and its thermal bath are quantized, so that $H_C$ is exactly conserved at every step. Another feature is that no statistical factor is necessary; the $i^{th}$ spin flips every time its resulting conjugate variable is $k_i \geq 0$. Although randomness can arise in the site selection, if the sites are chosen in an orderly fashion, then the dynamics is fully deterministic. Because the dynamics is governed by conservation of energy it is adiabatic, having a temperature that fluctuates with an average value given by $<T>=4J/[k_B \ln(1+4J/<k>)]$, where $<k>=<K/N>$ is the average kinetic energy per spin. Thus, the kinetic energy acts as an ideal Bose-Einstein gas that senses the true local temperature of the spin system.

Our model is similar to a kinetic-Ising model with competing dynamics [17], but with constraints on the dynamics that allow thermal equilibration. Previously used constraints include Kawasaki spin exchange, Glauber dynamics at fixed $T$, and the Creutz criterion for fixed energy. Some simulations have combined Kawasaki and Glauber dynamics [18], but with different temperatures for the different dynamics that results in non-equilibrium behavior. The fastest steps in our k-C model combine Kawasaki and Creutz dynamics, yielding microcanonical constraints that are appropriate for energy-conserving spin exchange. Intermediate steps involve single-spin-flip activation using the local average temperature, which lowers the free energy below that of the Ising model at $T=<T>$, consistent with the nanothermodynamic equilibrium of Landau theory [19]. The slowest steps change the size of dynamically-correlated regions, which yields a fully-open (generalized) ensemble from nanothermodynamics [20]. We call this combination of different constraints for different degrees of freedom a nanocanonical ensemble.

Figure 1 shows the temperature dependence of the average potential energy per spin, $<u>=<U>/N$, for various probabilities of spin-exchange ($p_{Exch}$) and single-spin-flip ($p_{Flip}$) dynamics. All of these results were obtained using the same computer program on the same $N$=12x12x12 lattice by simply changing the constraints on the dynamics. Constraints used for single-spin flips include a fixed-$T$ Glauber criterion $p_{Flip}(T)$, the fixed-energy Creutz criterion $p_{Flip}(H_C)$, or a fixed-$H_C$ and average-temperature criterion $p_{Flip}(H_C,<T>)$ (to be described below). Whereas spin exchange may have no constraints $p_{Exch}(-)$, the fixed-$H_C$ criterion $p_{Exch}(H_C)$, or both fixed-$H_C$ and average region-size constraints $p_{Exch}(H_C,<N_R>)$ (also to be described below). The flip-attempt probability $p_A$ is an adjustable parameter that controls the fraction of each type of dynamics, so that $1-p_A$ is the attempt probability for exchange, and $(1-p_A)/p_A$ gives the average number of exchange steps between each flip attempt. Note that $p_A*p_{Flip}(T)$ forms the familiar expression of an attempt frequency times a Boltzmann activation rate. The usual Ising model is obtained using $p_A$=1 with $p_{Flip}(T)$ (solid curve). The square symbols come from adding an average of 99 unconstrained exchange steps ($p_A$=0.01) between each flip attempt, 99*$p_{Exch}(-)$: $p_{Flip}(T)$. Although the spin-flip activation is at fixed-$T$, the absence of constraints on the spin exchange corresponds to infinite temperature, so that the competition between dynamics at such disparate temperatures yields stationary states with energies far above the Ising equilibrium. The Creutz model $p_{Flip}(H_C)$ (dashed curve) matches the Ising model, except near $<T>=T_C$ where finite-size effects give slight differences between the fixed-$H_C$ and fixed-$T$ simulations. However, in contrast to fixed-$T$ dynamics, there is no measurable change from the Creutz model by adding an average of 99 fixed-$H_C$ exchange steps between each flip attempt, 99*$p_{Exch}(H_C)$: $p_{Flip}(H_C)$ (triangles). Thus, the same steady-state behavior is attained no matter how rarely the system couples to its spin bath, provided the dynamics is always constrained by constant $H_C$. By analogy, the thermal equilibrium will be attained no matter how rarely the system couples to its thermal bath, provided the dynamics is also constrained by constant $H_C$. Because the Creutz model is deterministic it involves no transfer of heat. We argue that for the normal state of bulk materials, a Boltzmann-weighted step should be added to the simulation to yield thermal equilibrium behavior. The circular symbols in Fig. 1 are from our k-C model, which includes a Boltzmann probability for the single-spin flips, $99*p_{Exch}(H_C,<N_R>):p_{Flip}(H_C,<T>)$.

We start the k-C simulation by subdividing the cubic 12x12x12 Ising lattice into equal-sized cube-shaped regions, so that the initial number of spins in each region may be $N_R$=1728, 216, 64, 27 or 8. Periodic boundary conditions are used on all external surfaces, with the same interaction strength between all nearest-neighbor spins, so that the regions are defined solely by their dynamics. Each MC step (MCS) involves four possible substeps. The substeps, listed in order of decreasing likelihood for $p_A$ <0.5, are: 1) spin exchange, 2) spin-flip attempt, 3) spin-flip activation, and 4) change in size of the correlated region. The sequence is begun by randomly choosing a spin and one of its nearest neighbors from the lattice. Then with probability $(1-p_A)$, the most likely substep is 1) spin exchange, which occurs with probability $p_{Exch}(H_C,<N_R>)$=1 if the spin and its neighbor are in the same region and if the resulting conjugate variables are non-negative; otherwise $p_{Exch}(H_C,<N_R>)$=0. Note that this substep is adiabatic, with no Boltzmann factor or explicit temperature dependence, but it differs from the deterministic Creutz model by the random choice of two sites and the constraint of spin exchange. With probability $p_A$, the sequence will instead progress to substep 2) where the spin attempts a single-spin flip. If the Creutz criterion $k_i \geq 0$ is met, then 3) spin-flip activation occurs with probability

$$p_{Flip}(H_C,<T>)=\exp[<m><u>/k_B<T>]. \qquad \text{Eq. (1)}$$

Here $m$ is the number of spin-exchange steps in the region since the last flip attempt in the region, and $u = \Sigma_j u_j /N_R$ (with the sum over all spins in the region) is the interaction energy per spin in that region. The averages $<\bullet>$ are taken over all attempts since the last spin-flip activation in the region, so that $<u>$ approaches the mean-field interaction energy per spin and $<m> \sim (1-p_A)$

/$p_A$. Note that <$u$> is usually negative, so that Eq. (1) implies an Orbach process [21]; otherwise we set $p_{Flip}(H_C,<T>)=1$ as in the Metropolis algorithm. Eventually, after what may be many attempts at low <$T$>, a successful spin flip may also 4) change the size of the region, subject to constraints that limit fragmentation. Although substep 4) is essential for finding the equilibrium distribution of region sizes, for size-dependent behavior at fixed-$N_R$ this substep is omitted.

The open circles in Fig. 1 show that the potential energy of the Creutz Hamiltonian is reduced by including the thermal step. This reduction in <$u$> comes from enhanced local-spin alignment due to correlated spin exchange, while the entropy remains high from finite size of the correlations. Indeed, Fig. 2 shows that simulations with fixed region sizes [99*$p_{Exch}(H_C,N_R)$: $p_{Flip}(H_C,<T>)$], which match the Creutz model in the limit $N_R \rightarrow N$, have much lower <$u$> as $N_R \rightarrow 8$. However, thermal equilibrium usually involves a balance between energy and entropy. Although subdividing the lattice reduces the entropy of the spins, it also increases <$T$>, which increases the entropy of the kinetic energy reservoir. Maximizing the combined entropy of the spins and their bath to find the true thermal equilibrium yields a distribution of region sizes, as shown by the open circles in Figs. 1 and 2.

The key assumption in Eq. (1) is that spin-flip activation involves the average <$u$>, not the microscopic $H_C$ used for spin exchange. We argue that since the average temperature comes from the average kinetic energy, the activation energy should come from the average potential energy. In other words, although microcanonical spin exchange involves microscopic variables, thermalizing spin flips should involve thermodynamic variables. Our assumption may be related to the dynamical-averaging methods used to evaluate time-dependent Hamiltonians in magnetic spin resonance [22], adapted to treat spin relaxation rather than resonance. Another analogy is that $p_{Exch}$ and $p_{Flip}$ may be related to spin-spin and spin-lattice dynamics, respectively. There are at least three distinct scenarios that can yield Eq. (1). a) For a spatial cluster containing an average of <$m$> spins (inside the region containing $N_R$ spins), with an average interaction energy of <$u$> per spin, the probability of simultaneously activating all of these spins into a transition state of zero interaction energy is $p_{Flip}(H_C,<T>)$. b) If a successful spin flip requires repeated activation during every step between attempts, with the probability $p_1=\exp[<u>/k_B<T>]$ for each step, then after $<m>$ steps the net spin-flip probability is $p_1^{<m>}=p_{Flip}(H_C,<T>)$. c) Because $<m>$ is the average number of exchange steps between each flip attempt, then the effective coupling strength to the thermal bath is $J^*=J/<m>$, yielding an effective temperature of $<T^*> = <T>/<m>$ and an average spin-flip probability of $\exp[<u>/k_B<T^*>] = p_{Flip}(H_C,<T>)$. In any case, since each region is averaged over many microcanonical configurations before each attempt to couple to its environment, much of the randomness in the local properties is removed, similar to motional narrowing in resonance phenomena. This yields an average interaction between average degrees of freedom at average positions, justifying the use of a ferromagnetic lattice model for the slow dynamics and equilibrium properties of complex systems. Thus, like the universality near critical temperatures where local behavior is averaged over long distances, our simplistic model yields realistic behavior over a broad range of temperatures by averaging over long times.

Figure 3 is a histogram of some dynamical properties of the k-C model, presented in a manner that roughly mimics a log-log plot of frequency-dependent susceptibility. The solid symbols show the normalized distribution of spin-flip attempts. Thus, instead of the single attempt frequency of many models, the constant probability $p_A$ yields a broadened peak at high frequencies that is independent of $<T>$, which we associate with the boson peak in supercooled liquids. The open symbols show the relative distribution of successful spin flips, which shifts sharply towards lower frequencies with decreasing $<T>$, similar to the alpha peak in supercooled liquids. The solid curves in Fig. 3 come from fitting the spin-flip distribution to the Cole-Davidson formula for non-exponential response. Although the peak width is similar to that of a single-exponential response, these are histograms not susceptibilities, so that the entire width is due to thermal fluctuations. Also note that because the distribution is due to thermal fluctuations, not fixed energy barriers, the widths are effectively independent of $<T>$, yielding approximate time-temperature superposition. Figure 4 shows the most probable number of MC steps between each successful spin flip as a function of $1/\langle T \rangle$. The solid curves come from fitting these data to the Vogel-Tammann-Fulcher law. The inset shows that the resulting range of fragility parameters is similar to those found in many supercooled liquids.

In summary, we start with the Creutz model for a finite thermal bath and system of spins, then incorporate constraints that add realism to the dynamics. The constraints of conservation of energy and alignment are consistent with the picture of quantum degrees of freedom in correlated regions; as is the spatial averaging that yields indistinguishability of the spins in each region, and the temporal averaging that decreases the uncertainty in the energy with increasing time. We also incorporate several features from nanothermodynamics. One feature is that each region is in thermal contact with an ensemble of similar regions, not an infinite external bath, yielding a self-consistent internal temperature. Another feature is the use of microcanonical constraints for fast degrees of freedom that do not have time to couple to their environment, and the generalized ensemble for slow degrees of freedom that must couple fully to the surrounding sample without fixed-size constraints. Finally, because our k-C model is based on maximum entropy, not fixed temperature, it can simulate non-equilibrium conditions such as those in nonresonant spectral hole burning [23], dynamic specific heat [24], or other situations involving thermal gradients. These features of nanothermodynamics are crucial for simulating the properties of nanoscale systems, such as supercooled liquids with dynamical heterogeneity.

We thank A. K. Chizmeshya, K. C. Dixon, R. Richert, K. E. Schmidt, and G. H. Wolf for their contributions to this research. RVC thanks the Humboldt Foundation and A. Loidl for their support and hospitality while part of this work was done at the University of Augsburg Germany.

Figure captions

Fig. 1. Temperature dependence of the average potential energy per spin, $<u>=<U>/N$, for spin-exchange ($p_{Exch}$) and single-spin-flip ($p_{Flip}$) dynamics with various constraints. The temperatures are normalized by the critical temperature for the Ising model on a simple-cubic lattice ($T_C = 4.5115 J/k_B$). The solid curve comes from fixed-$T$ Glauber dynamics, $p_{Flip}(T)$. The dashed curve comes from fixed-energy Creutz dynamics $p_{Flip}(H_C)$. The symbols come from Kawasaki and Glauber dynamics □-99*$p_{Exch}$(-):$p_{Flip}(T)$, Kawasaki and Creutz dynamics Δ-99*$p_{Exch}(H_C)$:$p_{Flip}(H_C)$, and our k-C model ○-99*$p_{Exch}(H_C,<N_R>)$: $p_{Flip}(H_C,<T>)$.  Error bars are smaller than the symbol size, unless shown.

Fig. 2. Average temperature (positive values) and average interaction energy per spin (negative values) as a function of the number of spins in each correlated region. The four sets of symbols (from upper to lower for both quantities) have fixed total energies of $H_C/J$ = 41,600, 25,600, 16,000, and 8000. The solid symbols come from simulations with fixed region size 99*$p_{Exch}(H_C,N_R)$:$p_{Flip}(H_C,<T>)$; while the open circles come from the k-C model with fluctuating region size 99*$p_{Exch}(H_C,<N_R>)$:$p_{Flip}(H_C,<T>)$. Error bars are smaller than the symbol size, unless shown.

Fig. 3. Log-log histogram of spin-flip attempts (closed symbols) and successful spin flips (open symbols) at four average temperatures: $<T>/T_C$=1.17-□, 1.26-○, 1.55-Δ, 2.94-∇. The abscissa is obtained from the inverse number of MC steps since the last attempt (spin flip) rounded to the nearest power of 2. The ordinate gives the relative probability of each type of dynamics; obtained by counting the number of events in each bin, dividing by the number of steps in the bin, and multiplying by the number of MC steps between each event. Each set of data is normalized by the maximum attempt probability. The solid curves are fits to the open symbols using the Cole-Davidson function [1]. Inset: logarithm of the normalized probability of each value of $k_i$ as a function of $k_i$. The lines are fits to the data showing the Boltzmann probability for the kinetic energies, with a slope that gives $-4J/k_B<T>$.

Fig. 4. Angell plot of the most probable number of MC steps between each single-spin flip (MCS$_{peak}$) as a function of inverse average temperature for $p_A$ =0.005-○, 0.01-△, and 0.02-▽. Each curve is shifted vertically to give $10^{-14}$ as $T_g/<T>\rightarrow 0$, and normalized horizontally to give $10^2$ as $T_g/<T>\rightarrow 1$, which defines the glass transition temperature and is given approximately by $T_g=T_C/(1.0+17p_A)$. The solid curves are fits to the data using the Vogel-Tammann-Fulcher law [1]. The inset shows the fragility parameter, $M=d\ln(MCS_{peak})/d(T_g/<T>)|_{(T_g/<T_g>)}$, as a function of $p_A$. The solid line shows that $M=61+4700p_A$ over this range of $p_A$.

<ск>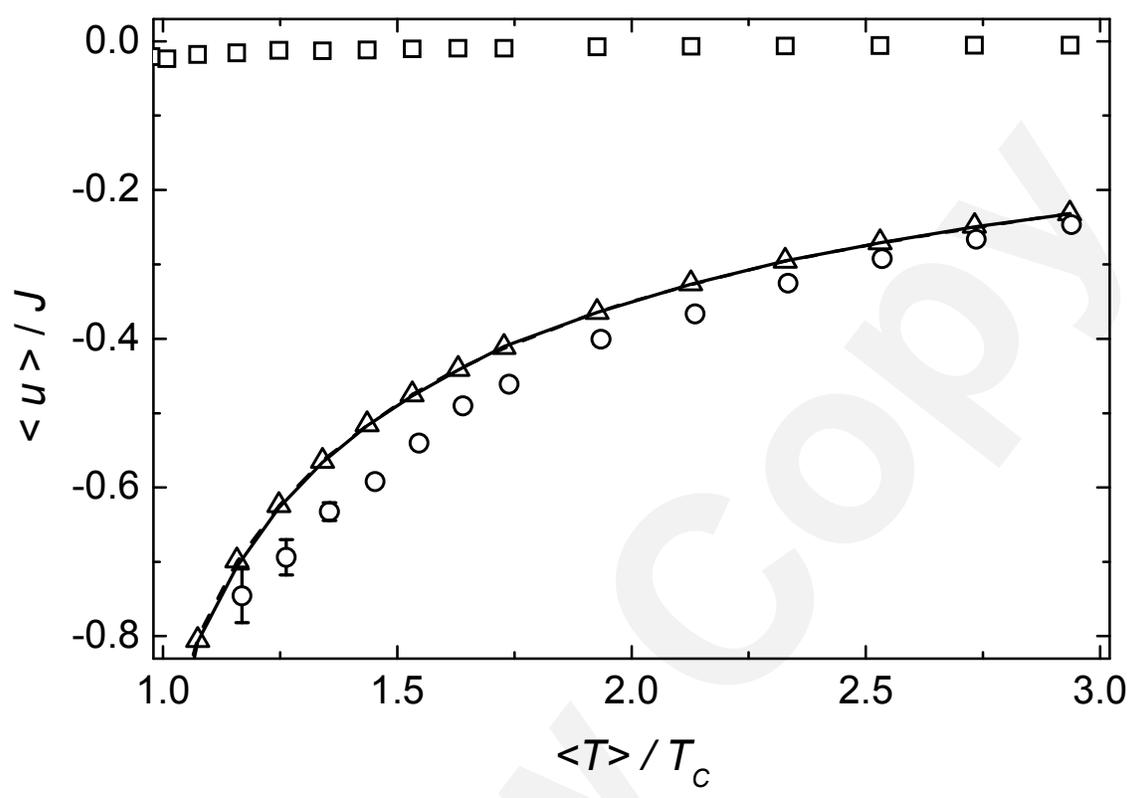

Figure 1

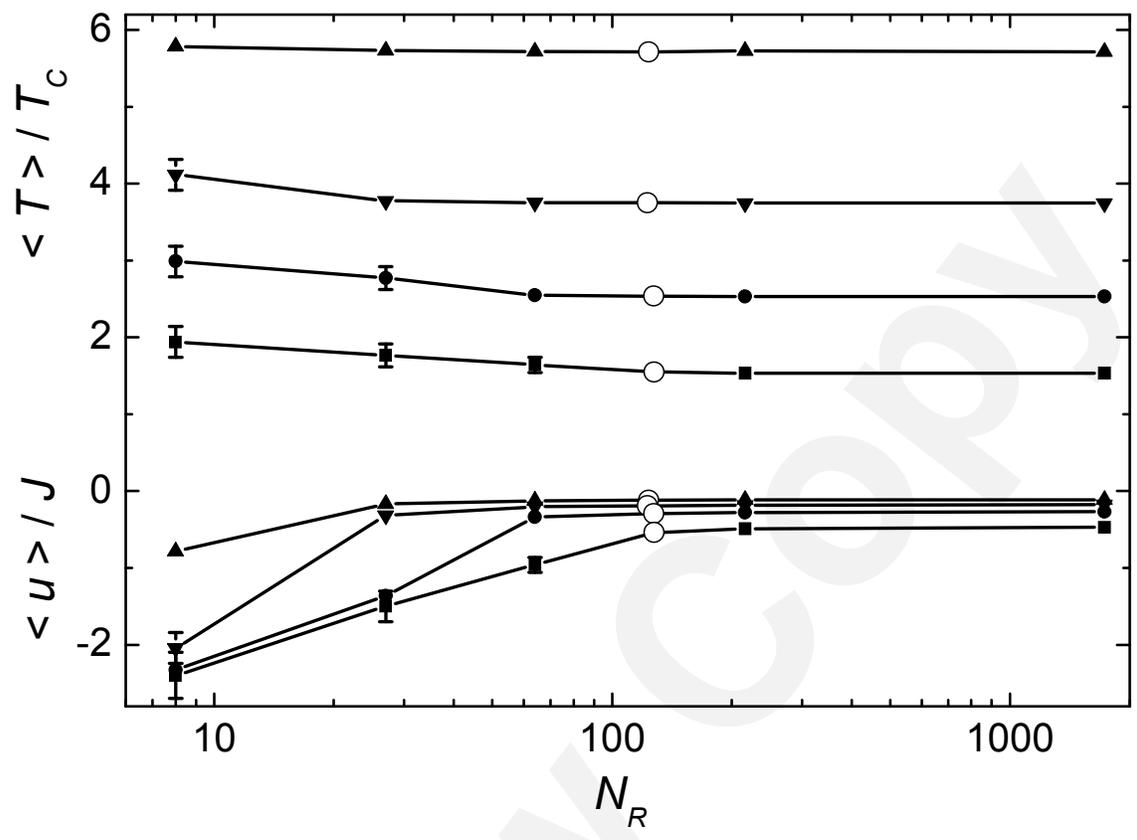

Figure 2

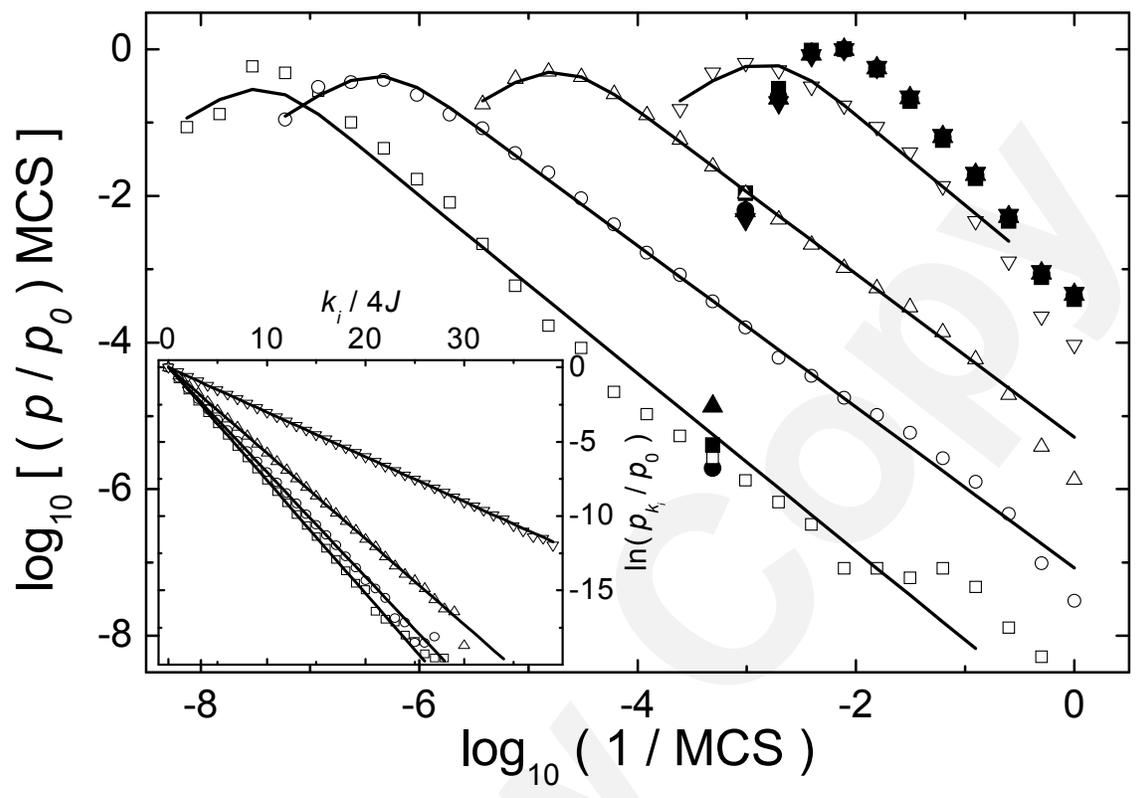

Figure 3

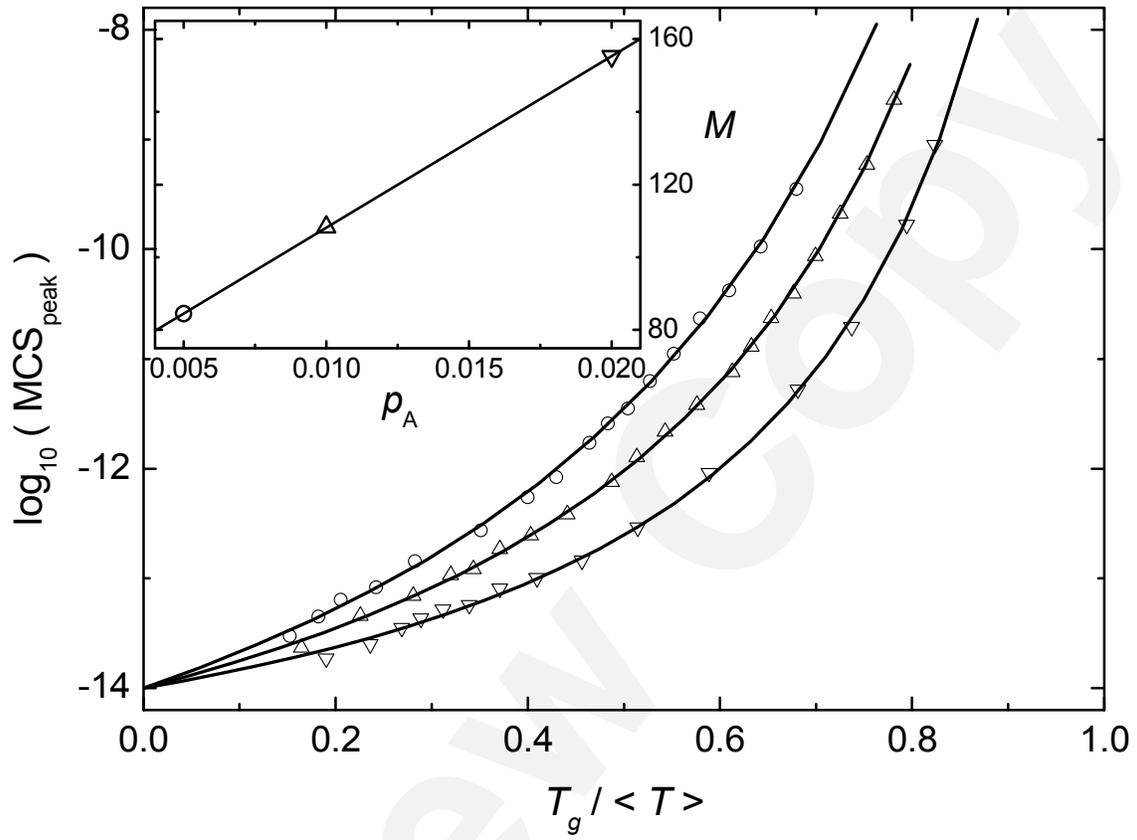

Figure 4